\documentclass[aps,prl,groupedaddress,a4paper,twocolumn]{revtex4}
\usepackage{graphicx}

\begin{document}

\title{Red solitons: evidence of spatio--temporal instability
  in $\chi^{(2)}$ spatial-soliton dynamics}

\author{Stefano Minardi}
\altaffiliation[Permanent address:] { Institute of Photonic
Sciences, Jordi Girona 29, NEXUS II, E-08034 Barcelona, Spain}
\author{Giovanni Blasi}
\author{Paolo Di Trapani}
\affiliation{INFM and Dipartimento di Scienze Chimiche, Fisice e Matematiche,
  Universit\`a dell'Insubria, Via Valleggio 11, I-22100 Como, Italy}

\author{Arunas Varanavi\v{c}ius}
\author{Gintaras Valiulis}
\author{Audrius Ber\v{z}anskis}
\author{Algis Piskarskas}
\affiliation{Department of Quantum Electronics, Vilnius University, 9
  Sauletekio, building 3, LT-2040 Vilnius, Lithuania}

\date{\today}

\begin{abstract}
  In $\chi^{(2)}$ three-wave mixing, the noise-seeded spatio--temporal
  modulational instability has a dramatic impact on the spatial-soliton
  dynamics, leading to the counterintuitive observation of a soliton with
  no apparent participation of the high-frequency field in the process.
\end{abstract}


\maketitle

\begin{figure}[hbt]
  \includegraphics[width=8cm]{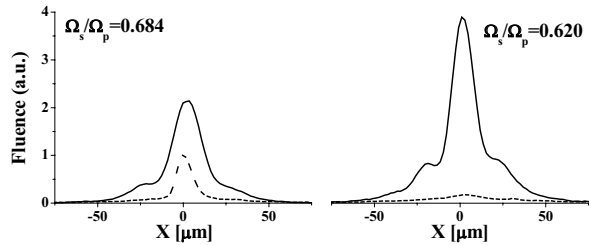}
  \caption{Transverse fluence distribution for the output signal and pump beams
    for OPA excitation and two different frequency-tuning conditions (a and b),
    with the same pump-pulse energy, $E_p=2.1\,\mu$J.}
  \label{fig:1}
\end{figure}

\begin{figure*}[hbt]
  \includegraphics[width=17cm]{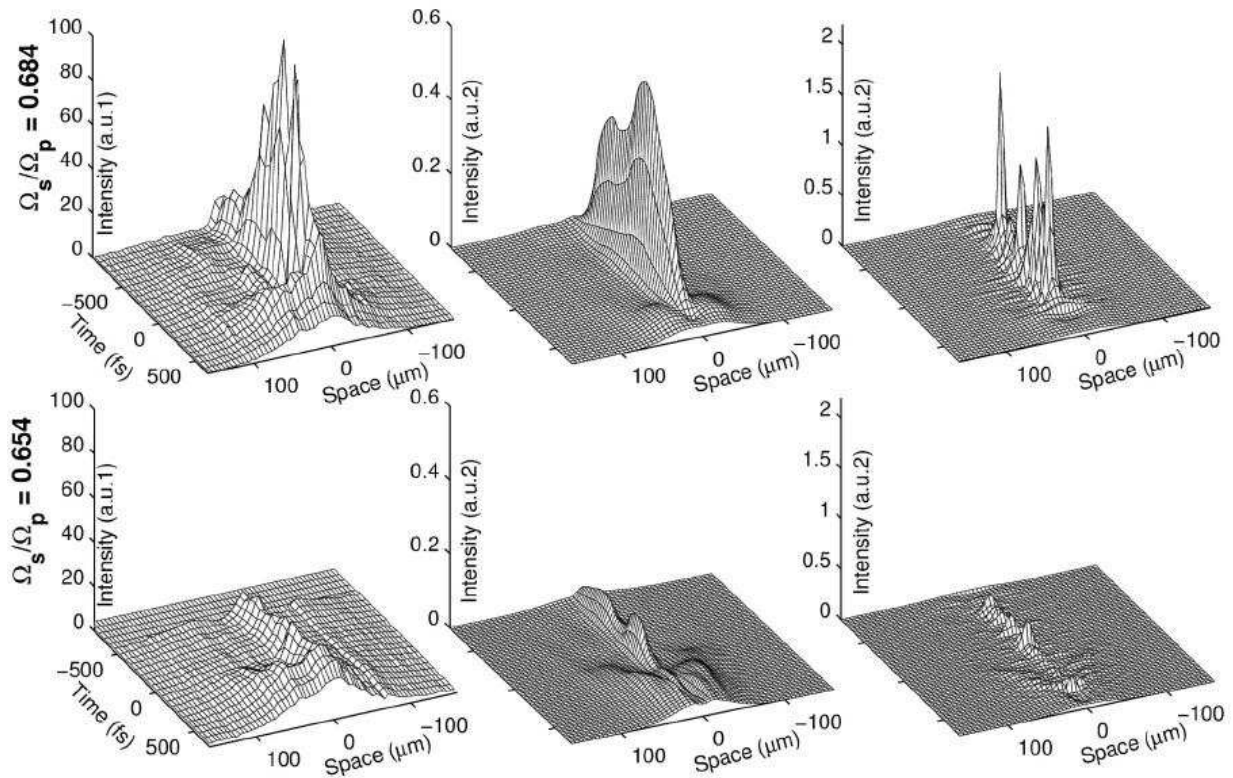}
  \caption{Spatio-temporal intensity maps of the output pump-wave packet for
    OPA excitation and two different frequency-tuning conditions (top and
    bottom); left: experimental data; center: smoothed numerical data; right:
    high resolution numerical data. All data refer to the same input-pulse
    energy.}
  \label{fig:2}
\end{figure*}

\begin{figure}[hbt]
  \includegraphics[width=8cm]{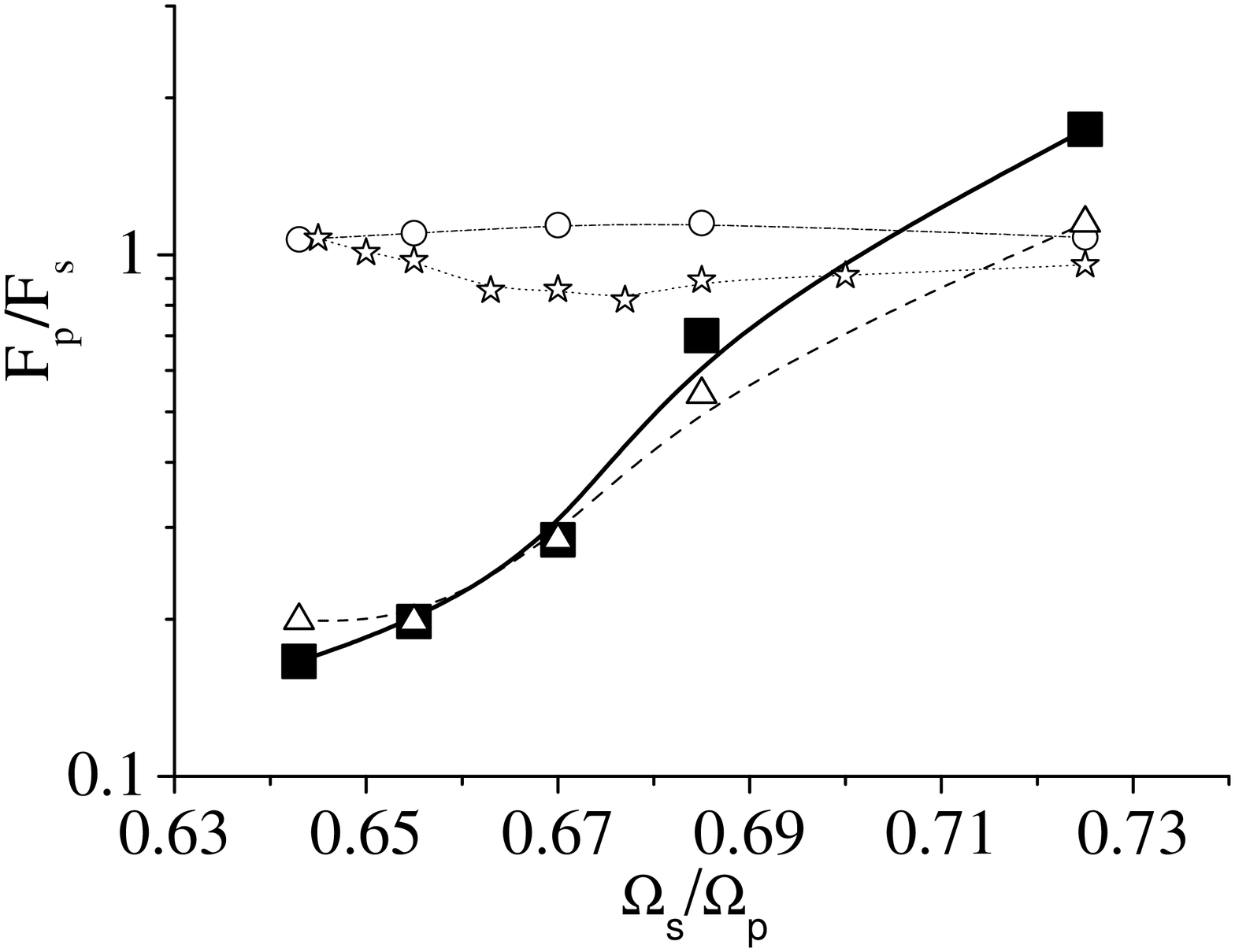}
  \caption{Frequency-tuning dependence of the pump-to-signal (peak) fluence
    ratio, $F_p/F_s$, for OPA excitation. Squares: experimental; Triangles: 3D+1
    spatio--temporal model; Circles: 2D+1 spatial model; Stars: 1D+1 temporal
    model. Fluences from the 3D+1 model are calculated by temporal integration.
    All data refer to the same input (peak) intensity.}
  \label{fig:3}
    \end{figure}

Optical spatial solitons \cite{sp_sol} are ideally monochromatic
light beams whose linear diffractive spreading is compensated by a
suitable non-linear phase shift. In $\chi^{(2)}$ three-wave
mixing, multi-color solitary beams are sustained by energy
exchange among the three waves, whose phase-velocity and
wavefront-curvature mismatch give rise to an intensity-dependent
phase accumulation \cite{Steg_rw}. Since their first observation
in a second-harmonic generation (SHG) scheme \cite{SHsol},
$\chi^{(2)}$ spatial solitons have been investigated in a number
of different experimental conditions, \emph{i.e.}, in up and down
conversion, in bulk, wave-guides and periodically-poled crystals,
with gaussian and vortex-type pumps; their unique steering,
addressing and switching properties have been studied in great
detail for potential applications in all-optical interconnects
(see reference \cite{reviewSHG} for an updated review on the
topic).
\newline\indent
With the exception of two recent experiments, where fairly short
pulses and long samples were used \cite{NLX, STWG}, all the above
mentioned phenomenology has been investigated in conditions in
which material chromatic dispersion was assumed to play a
negligible role. However, no matter how long the input pulses are
(\emph{i.e.}, how narrow the input pulse spectra are), one should
expect appreciable frequency broadening to occur, due the effect
of the noise-seeded modulational instability (MI) \cite{MI}. In
the spatial domain, sizeable MI impact has already been shown to
cause spatial filamentation of the SHG-generated temporal soliton,
for operation just above threshold \cite{Wise_MI}. By analogy, one
would expect temporal pulse fragmentation to affect the
spatial-soliton regime. This process is, however, difficult to
detect directly since it would require ultra-high temporal
resolution in single-shot acquisition mode.
\newline\indent
In this paper we present the first evidence for the effect of
noise-seeded spatio--temporal (ST) MI on $\chi^{(2)}$
spatial-soliton dynamics, which we predict to cause a temporal
breakup on the 10-fs scale. We claim ST MI to be the mechanism
that explains our counter-intuitive observation of a spatial
soliton for which self-trapping only occurs between the
low-frequency waves (\emph{e.g.}, the signal and idler), with no
apparent participation of the high-frequency wave (\emph{e.g.},
the pump). This is the reason we use the term ``red soliton'' to
indicate such a process even though, strictly speaking, no
solitons are formed.
\newline\indent
Our first experiment was performed in the regime of optical
parametric amplification (OPA) of the vacuum-state fluctuations,
with the usual scheme assumed to lead to spatial-soliton formation
\cite{sp_sol_W}: a pump pulse (527\,nm, 1.3\,ps, 32\,$\mu$m spot
FWHM) was launched on the input face of a 15\,mm-long lithium
triborate (LBO) crystal, operated in non-critical, type-I phase
matching. For the given pump frequency ($\Omega_p$), we tuned the
signal and idler frequencies ($\Omega_s$ and $\Omega_i$) by
changing the crystal temperature. Since the chromatic dispersion
(\emph{i.e.}, the group-velocity mismatch GVM and the
group-velocity dispersion GVD) experiences large changes with the
tuning parameter $\Omega=\Omega_s/\Omega_p$, the setup allowed
easy monitoring of the effects of dispersion on the non-linear
dynamics. We should mention that here that the GVM and GVD lengths
are respectively 2--4 and $10^2$--$10^3$ times larger than the
crystal length, across the whole tuning range.
\newline\indent
Two different diagnostics were realized. The first is a
traditional, CCD-based, beam-profile detection, which allowed
recording of the fluence distribution of the pump and signal waves
at the crystal output face. Note that the idler was not recorded,
its photon energy being too low for the silicon detector. The
second setup provided three-dimensional (\emph{i.e.}, ST) mapping
of the pump intensity profile, taken at the same crystal-output
plane. The scheme is based on an ultrafast non-linear gating
technique \cite{Lantz_gate} realized by mixing the unknown wave
packet (WP) with a 200-fs plane pulse generated by nonlinear
compression of a portion of the same pump-laser pulse. The
achieved temporal resolution is here of 200\,fs.
\newline\indent
Figure~\ref{fig:1} shows the fluence profiles along the beam diameter of the
pump (full line) and of the signal field (dashed line) just above threshold for
soliton formation.
Figure~\ref{fig:1}a refers to a crystal temperature $T=115^\circ$ (which
corresponds to $\Omega=0.684$ and $\lambda_s=770\,$nm), close to the edge of
the tuning curve; Fig.~\ref{fig:1}b refers to $T=140^\circ$ (and thus
$\Omega=0.620$ and $\lambda_{s}=850$), \emph{i.e.} much closer to degeneracy.
By looking at the result in Fig.~\ref{fig:1}a, one would reasonably affirm that
spatial solitons are formed. In fact, mutually trapped pump and signal beams of
comparable peak fluence appear. The pump FWHM beam diameter ($15\,\mu$m) is
five times smaller than the value obtained in the linear propagation regime. In
contrast, Fig.~\ref{fig:1}b contains quite a puzzling result. In fact, while
the signal keeps its narrow size, the focused pump beam almost disappears into
the weak background field. The background profile perfectly matches the
pump-beam pedestal in Fig.~\ref{fig:1}a. Note the growth of the signal peak
fluence from Fig.~\ref{fig:1}a to \ref{fig:1}b, its value being 20 times larger
than that of the pump in the second case. By taking a crystal of twice the
length we obtained very similar phenomenology, the diameters of the output
self-focused beams being only 30\% larger than in the previous case.
\newline\indent
A deeper insight into this rather counterintuitive observation is given by the
results presented in Fig.~\ref{fig:2} (left), where the experimental ST
intensity distribution of the pump WP is plotted for two different $\Omega$.
The results show that: (i) the trailing edge of the pulse (about
200\,fs) does not experience spatial trapping owing to its low
input intensity. This portion accounts for the pedestal and
diffracted background in Figs.~\ref{fig:1}a and \ref{fig:1}b
respectively. Note the slight temporal asymmetry of the entire ST
distribution, caused by GVM. (ii) The central and leading
portions, in contrast, always give rise to a spatially
self-trapped profile. Its peak intensity, however, decreases
dramatically as the tuning condition approaches degeneracy, which
explains why the trapping seems to disappear in the
time-integrated profiles. Time-resolved measurements closer to
degeneracy were prevented by the low sensitivity of our CCD video
trigger.
\newline\indent
In order to provide a more accurate description of the underlying
process, we performed high-resolution (4-fs step) numerical
experiments by solving the three-wave parametric equations
\cite{reviewSHG} within the framework of a 3D+1, ST model with
radial symmetry and temporal dispersion up to third order.
Figure~\ref{fig:2} (center) shows the calculated pump profiles for
conditions identical to those in Fig.~\ref{fig:2} (left). Here the
resolution of the numerical-data presentation is reduced to
200\,fs in order to match the resolution of the experimental
diagnostic. The excellent agreement between experiment and
calculation proves that the key features of the observed
phenomenon are determined by the $\chi^{(2)}$ process only.
Importantly, the full resolution plots, presented in
Fig.~\ref{fig:2} (right), indicate the occurrence of a temporal
breakup on the 10\,fs scale (the corresponding autocorrelation
shows a 25\,fs FWHM peak). The same structure occurs on the signal
and idler waves, which appear spatially and temporally
self-trapped with the pump. As to the dependence on $\Omega$, the
idler behaves in the same way as the signal wave.
\newline\indent
Evidence for the fundamental role played by the dimensionality in the
non-linear process is given by the plots in Fig.~\ref{fig:3}, where results of
the experiment, of the spatio--temporal, spatial and temporal models are
compared.
The results indicate that while the ST model describes the experiment fairly
well, both the monochromatic and plane-wave approximations completely fail to
reproduce the large variation of the fluence ratio with the tuning parameter,
giving an almost constant value (about unity) over the entire tuning range.
This unequivocally proves that the red solitons appear as a genuine ST effect.
Note that, since the temporal dynamics occurs on a time scale much shorter than
that of the pulse envelope, one should not expect any appreciable effect of the
input-pulse duration on the observed phenomenology. Direct numerical test with
5\,ps long pulses did not produce any detectable deviation from the results
shown in Fig.~\ref{fig:3}.
\newline\indent
Our claim is that the red solitons appear as a consequence of the
ST MI. In justifying this, we should first note that no theory is
yet available concerning the instability of an intense, CW and
focused pump beam. Therefore we must rely, in first approximation,
on the predictions valid for the CW and plane-wave regime. In this
case, quantum-noise parametric amplification has already been
shown to appear as a branch of the MI process \cite{Par_MI}. When
the ST features of the noise are accounted for, one should not
expect spatial and temporal breakup to occur as independent
processes. In fact, \emph{via} the phase-matching constraint, the
parametric gain (or, equivalently, the MI) couples space and time
by establishing angular dispersion and X-shaped ST spectra in the
amplified modes \cite{OPA_XMI}. We expect that the same key action
of the gain should play a role in our experiment as well. In fact,
we detected appreciable angular dispersion in the generated signal
wave. Moreover, even if the temporal breakup in Fig.~\ref{fig:2}
(right) appears similar to what might be expected by a purely
temporal MI, the numerical results in Fig.~\ref{fig:3} indicate
that the imbalance (\emph{i.e.} $F_p/F_s\ll 1$) is not retrievable
as long as spatial and temporal processes are considered
independently.

\begin{figure}[hbt]
  \includegraphics[width=8cm]{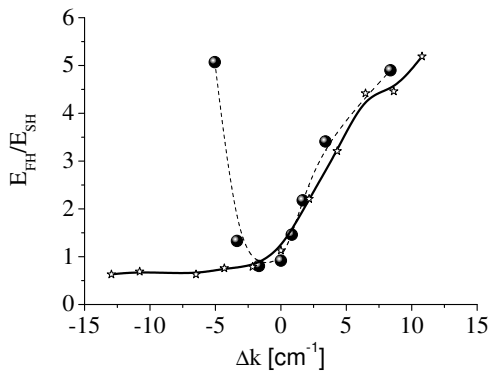}
    \caption{Phase-mismatch dependence of the ratio between the output FH and SH
    energies (experimental: balls), or intensities (CW model: stars), integrated within the FWHM diameter.
     All data are at taken at threshold.}
  \label{fig:4}
\end{figure}

In order to provide an explanation of the observed connection
between dimensionality, imbalance and dependence on $\Omega$ we
focus attention on the dramatic increase of the down-conversion
bandwidth close to degeneracy, which should cause a
down-conversion rate much larger than that for up-conversion and
thus the imbalance to appear. Note that a correlation between
bandwidths and imbalance must be expected as long as the signal
and idler field are not correlated. In fact, in case of
phase-conjugated fields, the expected drop in up-conversion
efficiency induced by the filtering action of the narrow
up-conversion bandwidth no longer occurs, mutual cancelation of
the opposite chirps taking place without losses in the
sum-frequency process \cite{chirpSF}. This consideration explains
why the 1D temporal model fails in retrieving the imbalance. In
fact, a phase conjugation between signal and idler is established
by the down conversion process. However, in the real 3D space,
this conjugation is quenched very rapidly owing to the fast phase
accumulation caused by diffraction and angular dispersion
\cite{Martinez86}. In contrast, in the plane-wave approximation,
it becomes unrealistically robust, due to the weaker effect of
GVD.
\newline\indent
With the aim of verifying the generality of the process we
performed a second experiment, where the spatial solitons are
excited \emph{via} SHG. The measurements, based on time-integrated
detection only, were performed by focusing down to a 45$\,\mu$m
spot (FWHM) a first-harmonic (FH) pulse (1.5\,ps, 1055\,nm) on the
input face of a 30\,mm long LBO crystal, operated in non-critical,
type-I phase matching. By changing the crystal temperature we
tuned the phase mismatch parameter, $\Delta
k=2k_\mathrm{FH}-k_\mathrm{SH}$, while keeping the frequency
tuning fixed at degeneracy ($\Omega=0.5$). The results in terms of
the ratio between the energy contents,
$E_\mathrm{FH}/E_\mathrm{SH}$, at threshold (\emph{both} FH and SH
beams trapped) are presented in Fig.~\ref{fig:4}, the $\Delta k<0$
region being limited close to phase matching by a rapid increase
of the threshold with $|\Delta k|$.
The few points, however, are sufficient to highlight the large
discrepancy occurring for $\Delta k<0$ between experiment and the
CW-model, which predicts solitons with the opposite imbalance with
respect to that measured.
\newline\indent
The fact that the CW model fails for negative, and not for
positive, phase mismatch must be related to the dominant SH
content expected in this parameter region, as long as spectral
broadening does not occur (see Fig.~\ref{fig:4}, full line). Due
to this, the dynamics that follows the SH generation must resemble
that of the OPA-excitation, where only the pump was launched at
the input. To this end we mention that, for $\Delta k\le0$, a
slight input-energy reduction caused the quenching of the trapping
only in the SH wave; this means an identical red-soliton
phenomenology as described for the OPA case. In comparing the two
excitation schemes, we expect that for the SHG case classical
noise should also contribute to the process. As to the impact of
the self-phase modulation, we mention that preliminary 3D+1
calculations performed in the absence of noise failed in
retrieving the imbalance. A second peculiarity of the SHG
excitation is that it leads to a true instability of the solitary
regime, a spatial soliton being expected in the absence of MI.
Again, since no theory has yet been developed for the MI of a bulk
solitary beam, we must rely on the study of the MI of the FH+SH
plane-and-monochromatic eigenmode for disclosing the mechanism of
the ST interplay. Indeed, in this case too the MI has been shown
to lead to the spontaneous formation of angular dispersion, its
gain curve having an X-type profile in the ST frequency domain
\cite{conicalSHG}.
\newline\indent
In conclusion, we have shown that conventional, time-integrated,
beam-profile measurements reveal the occurrence of a $\chi^{(2)}$
spatial soliton for which only the low-frequency signal and idler
waves appear to be self-trapped. The implementation of a
spatio--temporal, 3D mapping, technique has revealed that also the
high-frequency pump is trapped, but with an unexpectedly weak
intensity. Numerical calculations have highlighted the occurrence
of a temporal break up of all the three fields on a 10\,fs scale.
Moreover, they have shown that dispersion, diffraction and noise
are all necessary elements for describing the process. Relying
upon recent studies on the spatio--temporal instability of the
plane and monochromatic wave in OPA and SHG, we propose an
interpretation of the process as the manifestation of a
noise-seeded spatio--temporal modulational instability on the
$\chi^{(2)}$ spatial-soliton dynamics.
\newline\indent
The authors wish to acknowledge the technical assistance of
G.~Tamosauskas and D.~Mikalauskas, discussions with S.~Trillo and
C.~Conti, and support of the MIUR (Cofin01/FIRB02), UNESCO
UVO-ROSTE (875.586.2), EC CEBIOLA (ICA1-CT-2000-70027) contracts.

\end{document}